\documentclass[
reprint,
amsmath,
amssymb,
aps,
prapplied]{revtex4-2}

\usepackage{graphicx}
\usepackage{dcolumn}
\usepackage{bm}
\usepackage[colorlinks=true, allcolors=black]{hyperref}
\usepackage{physics}
\usepackage{nameref}
\usepackage{upgreek}

\usepackage{algorithm}
\usepackage{algpseudocode}
\usepackage[flushleft]{threeparttable} 
\usepackage{tabularx}
\usepackage{diagbox} 
\usepackage{mathtools}
\usepackage{booktabs}
\usepackage{longtable}
\usepackage{float}
\usepackage[shortlabels]{enumitem}
\usepackage[capitalize]{cleveref}
\usepackage{subfiles}

\begin{document}

\title{Parallel Tempering--Inspired Distributed Binary Optimization  \\with In-Memory Computing}

\author{Xiangyi~Zhang}
\affiliation{1QB Information Technologies (1QBit), Vancouver, British Columbia, Canada}

\author{Fabian~B\"ohm}
\affiliation{Hewlett Packard Labs, Hewlett Packard Enterprise, Milpitas, California, USA}

\author{Elisabetta~Valiante}
\affiliation{1QB Information Technologies (1QBit), Vancouver, British Columbia, Canada}

\author{Moslem~Noori}
\affiliation{1QB Information Technologies (1QBit), Vancouver, British Columbia, Canada}

\author{Thomas~Van~Vaerenbergh}
\affiliation{Hewlett Packard Labs, Hewlett Packard Enterprise, Milpitas, California, USA}

\author{Chan-Woo~Yang}
\affiliation{1QB Information Technologies (1QBit), Vancouver, British Columbia, Canada}

\author{Giacomo~Pedretti}
\affiliation{Hewlett Packard Labs, Hewlett Packard Enterprise, Milpitas, California, USA}

\author{Masoud~Mohseni}
\affiliation{Hewlett Packard Labs, Hewlett Packard Enterprise, Milpitas, California, USA}

\author{Raymond~Beausoleil}
\affiliation{Hewlett Packard Labs, Hewlett Packard Enterprise, Milpitas, California, USA}

\author{Ignacio~Rozada}
\thanks{{\vskip-10pt}{\hskip-9pt}Corresponding author: \href{mailto:ignacio.rozada@1qbit.com}{ignacio.rozada@1qbit.com}\\}
\affiliation{1QB Information Technologies (1QBit), Vancouver, British Columbia, Canada}

\date{\today}

\begin{abstract}
In-memory computing (IMC) has been shown to be a promising approach for solving binary optimization problems while significantly reducing energy and latency. Building on the advantages of parallel computation, we propose an IMC-compatible parallelism framework based on the physics-inspired parallel tempering (PT) algorithm, enabling cross-replica communication to improve the performance of IMC solvers. This framework enables an IMC solver not only to improve performance beyond what can be achieved through parallelization, but also affords greater flexibility for the search process with low hardware overhead. We justify that the framework can be applied to almost any IMC solver. We demonstrate the effectiveness of the framework for the Boolean satisfiability (SAT) problem, using the WalkSAT heuristic as a proxy for existing IMC solvers. The resulting PT-inspired cooperative WalkSAT (PTIC-WalkSAT) algorithm outperforms the standard WalkSAT heuristic in terms of the iterations-to-solution in 84.0\% of the tested problem instances and its na\"ive parallel variant (PA-WalkSAT) does so in 64.9\% of the instances, and with a higher success rate in the majority of instances. An estimate of the energy overhead of the PTIC framework for two hardware accelerator architectures indicates that in both cases the overhead of running the PTIC framework would be less than 1\% of the total energy required to run each accelerator.
\end{abstract}

                              
\maketitle

\section{Introduction}
In recent years, in-memory computing (IMC) has demonstrated promising results in solving binary optimization problems~\cite{bojnordi2016memristive, cai2019harnessing, pedretti2023zeroth, bhattacharya2024computing, sharma2023augmenting}, showing significant improvements in terms of reduced energy consumption, chip area, and time to solution. The small design and low-energy consumption of IMC hardware makes it a natural candidate for using multiple IMC cores to build highly parallelized optimization solvers. Parallelism is a well-established strategy for improving the performance of optimization solvers, from using the na\"ive  strategy of running parallel independent solver replicas with different initial solutions or random seeds to using more-advanced parallel heuristics~\cite{crainic2010parallel, harada2020parallel, silva2021quadratic}. 

Earlier work~\cite{crainic2010parallel, mcdonald2009parallel, crainic2004cooperative, polat2017parallel, jarvis2020cooperative} on cooperative parallelism heuristics, where multiple replicas of  solvers can communicate with each other, has shown that these heuristics can significantly outperform na\"ive parallelism strategies, at the cost of increased complexity and potentially higher energy consumption overhead. In this paper, we propose a simple algorithmic framework that leverages inter-replica communication across IMC solvers with a minimal increase in energy overhead. Our proposed method is directly related to the physics-inspired parallel tempering~(PT) algorithm, a technique used extensively in statistical physics and computational \mbox{optimization~\cite{hukushima1996exchange, geyer1991markov, moreno2003finding, zhu2020borealis}.} The PT algorithm runs multiple parallel replicas at different temperatures, periodically exchanging solutions between pairs of replicas at adjacent temperatures. The temperature parameter modulates each replica's exploration versus exploitation strategy. Higher-temperature replicas are better at exploring the solution space and can easily overcome local optima, whereas lower-temperature replicas are greedier and converge more quickly by closely following energy gradients~\cite{earl2005parallel}. The replica-exchange mechanism enables the transfer of information along the chain of replicas, combining the strengths of both approaches. The PT algorithm has been successfully applied to various optimization problems~\cite{zhu2020borealis, wang2009parallel, bagherbeik2020permutational, almeida4756904revisiting, kim2021physics}; for example, relevant to this work, a CPU-based Boolean satisfiability~(SAT) problem~\cite{cook2023complexity} solver using PT as its algorithmic engine was used by the team who won the 2016 MAXSAT competition~\cite{zhu2020borealis}. 

Our proposed framework is designed to run multiple IMC solvers in parallel, each representing a single replica in the PT algorithm framework. In-memory computing solvers must have a configurable parameter that represents the replica's temperature, and must be able to communicate with the other solvers to perform replica exchanges. In this context, the temperature refers to any parameter that has an effect on the balance between exploration and exploitation. Some examples of parameters that could be used as replica temperature analogues are the noise threshold in \mbox{WalkSAT} heuristics, the size of the tabu tenure in tabu heuristics, and the solution restart frequency. Almost every stochastic local search algorithm for binary optimization problems has similar tunable parameters, regardless of the hardware architecture, which highlights the framework's broad applicability. 

The framework is natively parallel, and generalizes concepts underlying the PT algorithm and other cooperative search algorithms; we thus refer to it as ``PT-inspired cooperative search'' or the ``PTIC framework'' for short. The PTIC framework can support a wide range of hardware accelerators and has a very low energy overhead, which is discussed in detail in Sec.~\ref{sec:hardware_implementation}. 

To demonstrate the effectiveness of the PTIC framework, we apply it to the SAT problem. The SAT problem was the first identified NP-complete problem~\cite{cook2023complexity, johnson1973approximation}. It is a problem of significant practical importance with many industrial applications, for example, in cryptography~\cite{mironov2006applications}, large-circuit verification~\cite{brand1993verification}, and artificial intelligence planning~\cite{castellini2003sat}. Because of the ubiquity of SAT applications~\cite{knuth2015art}, several IMC SAT solvers have recently been proposed~\cite{pedretti2023zeroth, bhattacharya2024computing, sharma2023augmenting, shim202430, xie202329}, most of them based on the WalkSAT heuristic~\cite{selman1993local}.

In this work, we use the PTIC framework to build a SAT solver where the replicas implement the WalkSAT heuristic~\cite{selman1993local}, as the WalkSAT heuristic is a performant heuristic that maps well to IMC hardware~\cite{xie202329}. Additionally, the WalkSAT heuristic is one of the most well-known and most studied SAT-solving heuristics, with fast, open source CPU versions freely available~\cite{walksatrepo}. By integrating the PTIC framework with the WalkSAT heuristic, we develop a distributed algorithm we call the \mbox{``PTIC-WalkSAT''} algorithm.

In summary, this study makes the following contributions:
\begin{enumerate}
    \item A parallelization framework, the PTIC framework, is proposed to improve the performance of IMC solvers for binary optimization with minimal energy consumption overhead.
    \item We apply the PTIC framework to build a SAT solver using the WalkSAT heuristic. The \mbox{PTIC-WalkSAT} algorithm is benchmarked to empirically prove its effectiveness with extensive computational experiments.
    \item A hardware energy model is proposed to evaluate the  energy consumption overhead of implementing the PTIC framework on two different IMC hardware architectures.
\end{enumerate}

This paper is organized as follows. Section~\ref{sec:bg_pre} introduces the SAT problem, the WalkSAT heuristic, and PT. In Sec.~\ref{sec:method}, we describe the SAT problem instances used for benchmarking and the PTIC-WalkSAT algorithm, as well as two benchmarking-related techniques: the evaluation metrics and the simulation environment. Section~\ref{sec:results} presents the full computational results. In Sec.~\ref{sec:discussion}, further analysis and discussion are provided. Section~\ref{sec:conclusion} gives a brief summary of our work, concluding the article. Appendices \ref{apx:internal_dynamics_ptic} and \ref{apx:full_benchmarking_results} provide additional information on the PTIC-WalkSAT algorithm and our full benchmarking results.

\section{Background and Preliminaries}
\label{sec:bg_pre}
Parallel tempering is a Markov chain Monte Carlo (MCMC) method~\cite{hukushima1996exchange,geyer1991markov} for obtaining the system configuration that has the lowest energy. In highly complex physical systems such as spin glasses, it is crucial to investigate the distribution of configurations at thermal equilibrium. This requires sampling from the system configurations associated with the lowest energy from the Boltzmann distribution. However, directly deriving samples from this distribution is a nontrivial task, which led to the development of MCMC methods. The core principle of MCMC methods is that, by one properly engineering the Markov chain, it will converge to the stationary probability distribution, which is the target distribution, regardless of the initial configuration. The step of obtaining a sample from the Markov chain based on the current state is referred to as performing an ``MC update''. Use of the Metropolis--Hastings (MH) algorithm~\cite{metropolis1953equation,hastings1970monte} to perform the MC update results in an MCMC method that serves this purpose. Theoretically, if the MH algorithm runs for an infinitely long time, it can reach all the configurations, including the global minimum energy configuration, which is the most stable state of the system. Unfortunately, in practice, the algorithm could become trapped among local minima for two reasons: the algorithm cannot execute indefinitely in practice, and the energy landscape of a system could be very rough at a low system temperature. A ``rough'' energy landscape is one where many local minima are separated by high energy barriers, making it very hard for the MH algorithm to reach the global minimum~\cite{hukushima1996exchange}.

To overcome energy barriers, PT runs the MH algorithm on replicas of the system, denoted by the list $\mathcal{R}= \{r_1, r_2,...,r_\kappa \}$, each of which is permanently associated with a specific index $i$ and is
characterized by a unique temperature from the list $\mathcal{T} = \{T_1, T_2, ..., T_\kappa \}$, with $T_1 < T_2, \ldots, < T_\kappa$, where $\kappa$ is the total number of replicas and temperatures, respectively. Let $x_i \in \{0,1\}^n$ denote the incumbent system configuration associated with replica $r_i$, where $n$ is the number of system variables. Even if replica exchanges are performed during the PT process, $r_i$ always refers to the $i$-th replica, maintaining its identity throughout the simulation. Similarly, $E_i$ refers to the energy of the incumbent configuration $x_i$. Mapping $\mathcal{E}$: $\{0,1\}^n \to  \mathbb{R}$ returns the energy value given a system configuration. In other words, we have $E_i = \mathcal{E}(x_i)$. After a predetermined number of MC updates at each temperature, configurations of the neighbouring replicas $r_{i-1}$ and $r_i$ are exchanged with probability 
\begin{equation}
    P(E_{i-1},T_{i-1} \leftrightarrow {E_i, T_i}) = \mathrm{min}(1, e^{\Delta\beta \Delta E}),
    \label{eq:replica_exchange}
\end{equation}
where $\Delta \beta = 1/T_i - 1/T_{i-1}$ and $\Delta E = E_i - E_{i-1}$, with $\Delta E$ representing the energy difference between the incumbent configurations in the replicas $r_{i-1}$ and $r_i$. 
The replica-exchange mechanism allows a lower-energy configuration in a higher-temperature replica to be transferred to a lower-temperature replica for intensification. Conversely, it enables a higher-energy configuration in a lower-temperature replica to be moved to a higher temperature for diversification. A ``sweep'' is defined as the period during which each replica independently updates all degrees (i.e., variables) of the system once. After completion of one or a chosen number of sweeps, a tentative replica exchange is performed. One run of PT could conduct many sweeps before convergence. In practice, a limit $S$ is put on the number of sweeps to prevent infinite loops from occurring. Other notation not defined elsewhere for PT is as follows:
\begin{itemize}[itemsep=0.1em]
    \item Rand(): a function used to generate a value in the interval $[0, 1]$ following some probability distribution (e.g., uniform).
    \item $Q$: the maximum number of steps (i.e., updates) performed within one sweep.
    \item $B$: the global variable used to track the best configuration, initialized with a starting random configuration.
    \item $E$: the energy value associated with the best configuration.
\end{itemize}
Algorithm~\ref{algo:pt_general} gives the pseudocode for the standard PT procedure.
\begin{algorithm}[H]
\caption{The parallel tempering framework}
\label{algo:pt_general}
\begin{algorithmic}[1]
\State Initialize global variables $\mathcal{R}$, $\mathcal{T}$, $B$, $E$;
\For{each replica $r_i$}
    \State Initialize the configuration $x_i$;
\EndFor
\For{sweep $s = 1,\ldots, S$}
    \For{each replica $r_i$}
        \State Perform a sweep to update the configuration $x_i$;
        \If{$E_i = 0$}
            \State \Return the configuration $x_i$;
        \ElsIf{$E_i < E$}
            \State $B = x_i$, $E = E_i$;
        \EndIf
    \EndFor
    \For{each pair of neighbouring replicas $(i, j = i+1)$}
        \State Compute the acceptance probability $P(E_{i},T_{i} \leftrightarrow {E_j, T_j})$ as per Eq.~(\ref{eq:replica_exchange});
        \If{Rand() $< P$}
            \State Swap the replicas $r_i$ and $r_j$;
        \EndIf
    \EndFor
\EndFor
\State \Return $B, E$
\end{algorithmic}
\end{algorithm}
\section{Methods}
\label{sec:method}

\subsection{The PTIC Framework}
\label{sec:ptic}

The PTIC framework uses an IMC binary optimization solver to perform  local updates to the configurations of the replicas. The ``local update'' can be seen as a step used to find a new configuration given the incumbent one. For binary optimization problems, any mapping \mbox{$\{0,1\}^n \to \{0,1\}^n$} can serve as the local update, where $n$ is the number of binary variables. However, there is an additional requirement for a mapping to be a valid local update: the mapping must be probabilistic, with the extent of this probabilistic behaviour governed by a noise parameter $\eta$. The parameter $\eta$ plays a crucial role in most heuristic solvers, as it helps them escape local minima. Consequently, this parameter is commonly found in many IMC solvers~\cite{cai2019harnessing, pedretti2023zeroth, bhattacharya2024computing}, which reflects the generality of satisfying the requirements of the local update. We use $\mathcal{L}_{\eta}$ to represent that the local update uses $\eta$, whereas $\mathcal{L}$ refers to the general case that does not specify the amount of noise.

\begin{figure}[H]
    \centering
\includegraphics[width=8.5cm]{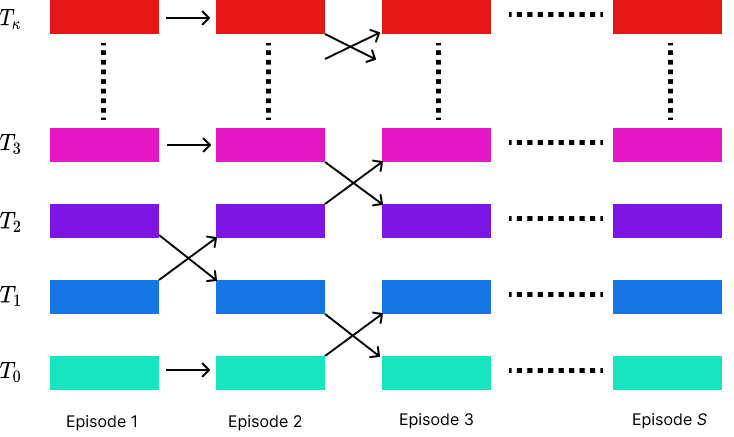}
    \caption{Replica-exchange visualization}
    \label{fig:pt_visualization}
\end{figure}

The default PT implementation uses the temperature of each replica to calculate the MH parameter used to accept or reject local updates: high-temperature replicas typically accept most configuration proposals, whereas low-temperature replicas tend to accept better proposals. In the PTIC framework, the temperature of each replica is reframed as the noise parameter $\eta$. With high values of $\eta$, the local update typically makes a random move, and with low values of $\eta$, it tends to move to a better configuration. The notion of a sweep in the PTIC framework is different from that of standard PT, as within a sweep of the PTIC framework, not every variable needs to be updated. Instead, we perform only a fixed number of local updates, which could leave some of the variables unchanged. To distinguish a sweep of the framework from a standard sweep, we call it an ``episode''.
At the end of each episode, tentative replica exchanges are performed. As illustrated in Fig.~\ref{fig:pt_visualization}, at the end of every episode, each pair of adjacent replicas can exchange their configurations with a probability defined in Eq.~(\ref{eq:replica_exchange}), where the energy $E_i$ refers to the number of violated clauses associated with the last configuration found by the replica $r_i$ before the episode ends.
A formal description of the PTIC framework is presented as Algorithm~\ref{algo:ptic_framework}. We use $\mathcal{P}$ to denote the binary optimization problem to be solved.
\begin{algorithm}[H]
\caption{The PTIC framework}
\label{algo:ptic_framework}
\begin{algorithmic}[1]
\State \textbf{Function} PTIC($\mathcal{P}, \mathcal{L}$):
\State Initialize $\mathcal{R}$, $\mathcal{T}$, $B$, $E$; 
\State Initialize a counter $s = 0$;
\While{$E >0$ and $s<S$}
\For{$i = 1,2, \ldots, \kappa$}
\State $x_i, E_i = \text{Episode}(x_i, Q, \mathcal{L}_{T_i})$; 
\If{$E_i = 0$}
    \State Return  $x_i$;
\ElsIf{$E_i < E$}
        \State $B = x_i$, $E= E_i$;
\EndIf
\EndFor
\For{$i=1,2,\ldots, \kappa -1 $}
 \State Compute the acceptance probability $P(E_{i},T_{i} \leftrightarrow {E_{i+1}, T_{i+1}})$ as per Eq.~(\ref{eq:replica_exchange});
        \If{Rand() $< P$}
            \State Swap the replicas $r_i$ and $r_j$;
        \EndIf
\EndFor
\State $s \gets s +1$;
\EndWhile
\State \Return $B, E$
\end{algorithmic}
\end{algorithm}
\subsection{Applying the PTIC Framework to the SAT Problem}

\subsubsection{The Boolean Satisfiability Problem}
The SAT problem can be formally described as follows. Given a propositional formula $\mathcal{F}$, expressed in the conjunctive normal form representation, defined by a set of variables $\mathcal{V}$ and a set of clauses $\mathcal{C}$, solving the SAT problem amounts to finding an assignment to the variables that renders all the clauses true. 
The following equation is an example of a SAT problem that is composed of four variables and four clauses:
\begin{eqnarray}
    \mathcal{F}(x_1, x_2, x_3, x_4) = (x_1 \lor \neg x_2 \lor x_4) \land (x_2 \lor \neg x_3)\nonumber \\ 
    \land (x_3 \lor x_4) \land (\neg x_1 \lor \neg x_3)
    \label{eq:sat_pre_eg}
\end{eqnarray}
 In general, an assignment to the SAT problem is a set of binary values that specifies whether a variable should have a value of 1 or 0. In our example, $x_1,x_2,x_3,x_4 = 0,0,0,0$ is an assignment to $\mathcal{F}(x_1, x_2, x_3, x_4)$ such that it makes true three of four clauses. An assignment that makes all the clauses true is considered a solution to the problem; for example, $x_1,x_2,x_3,x_4 = 0,1,0,1$ is a solution to the problem represented by the formula $\mathcal{F}$.

\subsubsection{The WalkSAT Heuristic}
The seminal paper by Selman, Kautz, Cohen, et al.~\cite{selman1993local} introduces WalkSAT heuristics, which generally begin with an initial configuration updated by flipping one variable at each iteration of a loop. One of the critical steps of WalkSAT heuristics is to decide which variable to flip, which depends on a scoring function. With different scoring functions, different versions of WalkSAT heuristics can be derived. In particular, in our study, we use the break value  to score a candidate variable as the default version implemented in the codebase~\cite{walksatrepo} uses the break value. The term ``break value'' refers to the number of newly violated clauses with respect to a variable, that is, the number of clauses that are satisfied in the current solution, but will become violated upon flipping of the variable. If not otherwise specified, ``WalkSAT heuristic'' refers to the version that uses the break value throughout the rest of this paper. Aside from the scoring function, the walk probability is a simple yet critical setting~\cite{selman1993local} because it prevents the heuristic from always performing a greedy move with respect to the break value, which could easily cause the heuristic to become trapped in local optima.
The pseudocode of the heuristic is presented in Algorithm~\ref{algo:walksat_b}.
\begin{algorithm}[H]
\caption{The WalkSAT heuristic}
\label{algo:walksat_b}
\begin{algorithmic}[1]
\State \textbf{Function} WalkSAT($\mathcal{C},x, Q, \eta$):
\State Initialize $q = 0$;
\While{any violated clause with respect to $x$ and $q<Q$}
    \State Randomly choose a violated clause; \label{walksat_b_step:random_clause}
    \State Initialize $l'$ to $0$ and $z'$ to a large number;
    \For{$l$ in the literals of the chosen clause}
        \State Compute the break value for $l$, denoted as $z(l)$;
        \If{$z' > z(l)$}
            \State $z' = z(l)$, $l' = l$;
        \EndIf
    \EndFor
    \If{Rand() $< \eta$}\label{algo_line:walk_prob_walksat_b}
        \State Randomly choose a variable from the clause to flip;
    \Else
        \State Choose the variable $\mathrm{abs}(l')$ to flip;
    \EndIf
    \State Update $x$ to align with the flipped variable; \label{walksat_b_step:flip_variable}
    \State $q \gets q+1$;
\EndWhile
\end{algorithmic}
\end{algorithm}
\subsubsection{The PTIC-WalkSAT Algorithm}

To embed the WalkSAT heuristic into the PTIC framework, the local update $\mathcal{L}$ that captures its main features must be identified. Since the key component of the WalkSAT heuristic is to decide the variable to flip and perform the flip at each iteration, we can define $\mathcal{L}_{\eta}$ as one iteration of the WalkSAT heuristic with the walk probability $\eta$. The local update is equivalent to executing lines \ref{walksat_b_step:random_clause}-- \ref{walksat_b_step:flip_variable} of Algorithm~\ref{algo:walksat_b}. The local update can also be referred to as a ``step'' or ``iteration'' in the context of the PTIC-WalkSAT algorithm. The local update is valid as it satisfies the two conditions defined in Sec.~\ref{sec:ptic}: flipping a variable is a mapping $\{0,1\}^n \to \{0,1\}^n$, and the walk probability provides randomness to the mapping. Regarding the notion of temperature, we reframe the walk probability as the temperature, because the walk probability manages the degree of greediness in performing the local update. In other words, in the context of the PTIC-WalkSAT algorithm, temperature and walk probability can be considered equivalent. Temperatures are bound as $0.0 < T_1,T_2,\dots, T_\kappa\leq 1.0$ because $\eta$ is a probability. 

\subsection{Hardware Implementation for the PTIC Framework} 
\label{sec:hardware_implementation}

\begin{figure*}[ht]
   \centering
\includegraphics[width=\linewidth]{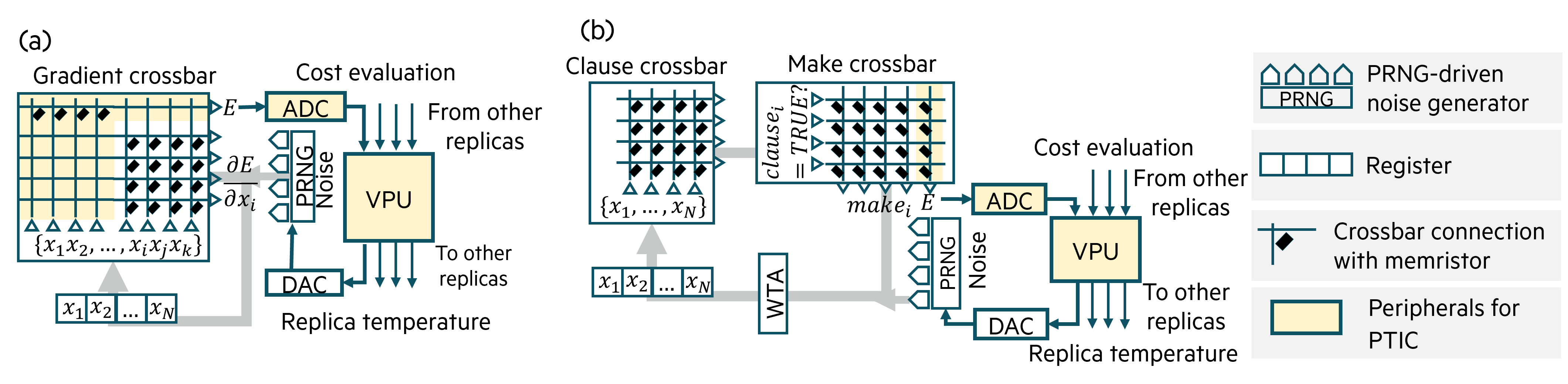}
   \caption{Exemplary hardware architectures for realizing the PTIC framework within (a) a PUBO and (b) a CAMSAT hardware accelerator for solving SAT problems. Peripherals added to support the PTIC framework are highlighted in yellow.}
   \label{fig:PTIC_HW}
\end{figure*}

The generic nature of the PTIC framework makes it suitable for implementation on a variety of hardware accelerator platforms. In what follows, we sketch possible implementations and consider the overhead of the PTIC framework for two example in-memory accelerators that can be used to solve SAT problems. Both accelerators use memristor crossbar arrays to perform massively parallel computation of gradients. The first accelerator, depicted in Fig.~\ref{fig:PTIC_HW}(a), implements a gradient-decent optimization of polynomial unconstrained binary optimization (PUBO) problems~\cite{hizzani2024memristor}. Boolean satisfiability problems can generally be expressed as PUBO problems by one translating their clauses into the generic polynomial cost function $E=\sum_{ij}A_{ij}x_ix_j+\sum_{kmn}B_{kmn}x_kx_mx_n + \cdots$, where $x_i=\{0,1\}$ are the Boolean variables and $A_{ij}$, $B_{kmn},\ldots$ are the linear coefficients for polynomials of a given order. The memristor crossbar is used for calculating gradients of the PUBO cost function, which is used to decide whether a specific variable needs to be flipped to minimize the number of unsatisfied clauses. To escape local minima, an analog noise signal is added to the gradients. In the accelerator, this is achieved with the use of an array of digital-to-analog converters (DAC), which is driven by a pseudorandom number generator. The magnitude of the generated analog noise signal is adjustable through a voltage signal provided by an additional DAC. On the basis of these noisy gradients, the register entries storing the Boolean variables are then flipped. For small-scale problems, such PUBO accelerators have been shown to be able to solve SAT problems while consuming only a few milliwatts of electrical power~\cite{hizzani2024memristor}. These accelerators can be adapted to the PTIC framework, where each replica is implemented as a new, separate PUBO accelerator. In this case, the temperature of each replica is given by the strength of the injected noise signal. To support the PTIC framework, a PUBO accelerator must be extended so as to calculate the overall cost for each replica. In addition, a central processing element is required that enables the replica-exchange process by comparing each pair of temperature-adjacent replicas.

To evaluate the PUBO cost function, determination of the overall cost for each replica is needed in order to calculate the acceptance probability in Eq.~(\ref{eq:replica_exchange}) during each replica exchange. In principle, the cost function can be evaluated inside the crossbar array at the expense of having a larger crossbar. As shown in Fig.~\ref{fig:PTIC_HW}(a), an additional crossbar column is used to sum the cost terms. Additional crossbar rows are also needed to calculate the cost of the individual clauses. Under the worst-case assumptions (i.e., when none of the cost terms are identical to the gradients of other clauses), twice the number of crossbar rows would be required. The output signal of this additional crossbar column is then sampled by an analog-to-digital converter (ADC) with a resolution of $\log_2(n_c)$, where $n_c$ represents the number of clauses. The PUBO cost for each replica is evaluated in the central processing element, which calculates the
acceptance probability, performs the replica exchange, and updates the temperature for each replica
. The temperature signal is then provided to the control DAC to adjust the magnitude of the analog noise signal within the replicas.

As the second example in-memory accelerator architecture, we consider a content-addressable memory--based (CAM) SAT solver (CAMSAT), depicted in Fig.~\ref{fig:PTIC_HW}(b). The CAMSAT accelerator implements a WalkSAT-like heuristic using crossbar arrays~\cite{pedretti2023zeroth}. This architecture consists of two crossbar arrays, where the first is used to evaluate violations of the SAT clauses. The output of the crossbar arrays uses a binary signal to indicate whether the clauses mapped to the respective row of the array are violated by the current variable configuration. This signal is passed to the second crossbar array, which is the transpose of the first array. The output of this second crossbar array indicates how many clauses become satisfied when  each of the variables is flipped (the ``make value'') during one iteration. A winner-takes-all circuit then selects the variable with the highest make value and flips its value. As with the PUBO architecture, an analog noise signal is added to aid the solver in escaping local minima. Adaptation of this architecture to the PTIC framework requires the calculation of the cost function, which is given by the number of violated clauses. This can be achieved within the second crossbar array with the use of an additional row that sums the input values. The output of this row is then digitized with an ADC with a $\log_2(n_c)$-bit resolution. As in the case of the PUBO accelerators, the replica temperature is given by the magnitude of the injected noise signal. The central processing unit used to perform the replica exchange is identical to that of the discussed PUBO architecture.

To estimate the potential overhead of adapting these accelerator architectures to the PTIC framework, we compare the energy consumption of an individual accelerator system with that of the additional peripheral components discussed. For the PUBO accelerator, we estimate that the overhead from the PUBO cost calculation would double the energy consumption of the crossbar array $E_{\text{xbar}}$ each time a replica exchange is performed, where the Boltzmann factors in Eq.~(\ref{eq:replica_exchange}) are evaluated. In addition, the ADC consumes $E_{\text{ADC}}$ at each evaluation. Of note, the cost calculation is required only during replica exchanges, such that the section of the crossbar array dedicated to the cost evaluation and the ADC are activated only every $Q$ iterations. The average energy overhead from the cost evaluation per solver iteration is thus $(E_{\text{xbar}} + E_{\text{ADC}})/Q$. In what follows, the subscripts ``stat'' and ``dyn'' represent the static energy and the dynamic energy, respectively, where static energy is consumed continuously, while dynamic energy is consumed only during a specific operation. The central processing unit can be realized with the use of vector processing units (VPU), which are similarly used in electronic artificial intelligence accelerators and digital signal processors. Such VPUs can perform parallel linear and nonlinear operations. The energy consumption of VPUs is estimated to be $E_{\text{VPU,stat}} + E_{\text{VPU,dyn}}/Q$ per iteration. As an example, we estimate the total PTIC framework's energy overhead for the PUBO architecture presented in Ref.~\cite{hizzani2024memristor}, designed to solve 20-variable \mbox{3-SAT} problems. 
In the referenced PUBO architecture, the total energy used is $1.3$ pJ per iteration. For the crossbar array, ADC, and VPU, we estimate the energy consumption per iteration on the basis of values obtained from the literature for the same technology node to be as follows: \mbox{$E_{\text{ADC, dyn}} = 1.5$ pJ~\cite{adc_survey},} $E_{\text{VPU,dyn}} = 2.2$ pJ, $E_{\text{VPU,stat}} = 0.25$ pJ~\cite{ankit2019puma}, and $E_{\text{xbar}} = 0.35$ pJ. Assuming that replica exchanges occur every $Q = 1000$ iterations, the PTIC framework's energy overhead constitutes $0.0086$ pJ per iteration, equivalent to $0.65$\% of the total energy consumption.

For the CAMSAT accelerator, the energy overhead can be estimated in a similar fashion. We assume that the overhead from the cost function evaluation is similar to the energy consumption of a single row in the second crossbar array. In contrast to cost function evaluations using the PUBO accelerator, each row is driven by the same input signals as the gradient calculation, such that we cannot assume it to be turned off outside replica exchanges. The overhead of the other peripheral components is identical to that of the PUBO accelerator, but must be adjusted for the longer iteration time of the CAMSAT accelerator. As an example, as we did in the case of the PUBO architecture, we estimate the energy overhead for a CAMSAT solver for 20-variable 3-SAT problems, where the average energy is approximately \mbox{$2.6$ pJ} per iteration. For the crossbar arrays' energy, we estimate $E_{\text{xbar}}=0.02$ pJ per iteration. Assuming that the replica exchange occurs every $Q=1000$ iterations, the PTIC framework's energy overhead constitutes $0.01965$ pJ per iteration, which is equivalent to $0.76$\% of the total energy consumption. 

\subsection{Benchmarking Methodology}

\subsubsection{The Problem Instances}
The problem instances used in our benchmarking experiments were chosen to be hard for \mbox{WalkSAT} algorithms. There are four groups of instances. \mbox{\textit{Group 1}} was generated from well-known semiprime factoring problems~\cite{bebel2019hardersat}. The number of variables in the instances ranges from 82 to 100, and the number of clauses ranges from 1177 to 2642. \textit{Group 2} comprises statistically hard random 4-SAT problems, each of which has a quietly planted solution~\cite{krzakala2012reweighted}. Each of the instances has 1000 clauses and 100 variables. To have statistically hard instances, the problems were generated at the computational phase transition threshold~\cite{mertens2006threshold}. A similar approach was used to generate hard 6-SAT and 7-SAT problems, denoted as \mbox{\textit{Group-3}} and \textit{Group-4}, respectively. Tables~\ref{tab:full_table_first_part} and~\ref{tab:full_table_second_part} give the key characteristics of the different problem classes. The base dataset included 558 problem instances, and we extracted 94 of them where the WalkSAT heuristic implemented in the codebase~\cite{walksatrepo} took more than $10^7$ iterations to find the solution. These 94 instances are considered to be the hard instances on which PTIC-WalkSAT was benchmarked. Tables~\ref{tab:full_table_first_part} and~\ref{tab:full_table_second_part} in Sec.~\ref{apx:full_benchmarking_results} of the appendix list these instances.
\begin{table}[H]
    \centering
    \begin{threeparttable}
    \caption{Summary of problem characteristics. ``K'' refers to the degree of the problems, and ``type'' refers to the class of problems in each group, namely semiprime factoring (SPF) instances, or random K-SAT. ``NC'' refers to the number of clauses; ``NH'' refers to the number of identified hard instances within each group; ``NI'' refers to the number of instances in a group; and ``NV'' refers to the number of variables.}
    \label{table:instance_summary}
    \begin{tabular}{|c|c|c|c|c|c|c|} 
        \hline
        \,\,\,Group\,\,\, & \,\,\,NV\,\,\, & \,\,\,NC\,\,\, & \,\,\,K\,\,\, & \,\,\,NI\,\,\, & \,\,\,Type\,\,\, & \,\,\,NH\,\,\, \\
        \hline
        1 & \, [82, 100]\tnote{a} \, & \, [1177, 2642]\tnote{a} \, & \, 11\tnote{b} \, & 258 & \, SPF \, & 4 \\
        2 & 100 & 1000 & 4 & 100 & 4-SAT & 1 \\
        3 & 50 & 2200 & 6 & 100 & 6-SAT & 35\\
        4 & 50 & 4500 & 7 & 100 & 7-SAT & 54\\
        \hline
    \end{tabular}
    \begin{tablenotes}
        \item[a] The value falls into the range
        \item[b] The median degree
    \end{tablenotes}
    \end{threeparttable}
\end{table}
\subsubsection{The Metrics}\label{sec:metric}
We use the iterations-to-solution metric, denoted as $ITS_{99}$, shown in Eq.~(\ref{eq:its}) to measure the performance of an algorithm for a given problem instance. The $ITS_{99}$ metric is defined as the number of iterations required to find a solution with 99\% certainty. To calculate $ITS_{99}$, $\gamma$ repeats are generated to gather statistics for each algorithm and problem instance, with the use of different seeds. The $ITS_{99}$ metric is given by
\begin{equation}
    ITS_{99} = \tau R_{99},
\label{eq:its}
\end{equation}
where $\tau$ is the upper bound of the number of iterations and $R_{99}$ stands for the number of repeats needed to find a solution with a probability of more than 0.99. Readers can refer to Ref.~\cite{aramon2019physics} for more details about the metric. 

In principle, we should use the metric ``energy to solution'' to evaluate the performance of an algorithm. However, since the energy overhead of the PTIC framework is minimal (see Sec.~\ref{sec:hardware_implementation} for details), we focus only on the energy consumed during each iteration (or local update). Given that the energy per iteration is consistent across all algorithms, we use $ITS_{99}$ as a proxy for energy to solution.

In particular, the total number of iterations performed by the PTIC-WalkSAT algorithm is defined as the total number of local updates performed by the successful replica multiplied by the number of replicas:
\begin{equation*}
    \kappa[Q(s-1) + q],
\end{equation*}
where $s$ is the number of executed episodes and $q$ is the number of steps that the successful replica performs in the last episode. 

When one is measuring the total number of iterations of other baseline algorithms, such as a parallelized algorithm that runs multiple independent replicas, the total number of iterations is the product of the number of replicas and the number of steps performed by the replica that has found the solution. The measurement is performed under an optimistic assumption that the replica that finds the solution is able to send an instant message that causes all the other replicas to halt.

To measure the difference between the PTIC-WalkSAT algorithm and a baseline, we use the percentage difference $\delta= [(I' - I)/I] \times 100\%$, where $I$ stands for the $ITS_{99}$ of the PTIC-WalkSAT algorithm and $I'$ is the $ITS_{99}$ of the baseline. We define $\delta \in [0.0,0.2)$ as a ``small improvement'', $\delta \in [0.2,0.8)$ as a ``medium-sized improvement'', and $\delta \in [0.8,\infty)$ as a ``significant improvement''. Finally, we use $\delta \in (-\infty,0.0)$ to mean that the baseline performs better than the PTIC-WalkSAT algorithm, meaning there has been a ``decline in improvement''.

Additionally, we define the notions of ``per-problem success rate'' and ``per-group success rate'', where the former refers to the average proportion of repeats that found the solution across the group (calculated with Eq.~(\ref{eq:per_problem_rate})) and the latter (calculated with Eq.~(\ref{eq:per_group_rate})) refers to the proportion of problems in each group where at least one repeat was able to find the solution. The per-problem success rate is given by
\begin{equation}
    \frac{1}{|\mathcal{I}|}\sum_{i \in \mathcal{I}}\frac{\gamma'_i}{\gamma} \times 100 \%, \label{eq:per_problem_rate}
\end{equation}
where $\gamma$ is the number of repeats for which a solver is run on a problem instance, $\gamma'_i$ is the number of successful repeats (meaning that a configuration has been found) for the instance $i$, and the set $\mathcal{I}$ stands for a group of instances of interest. The per-group success rate is given by
\begin{equation}
    \frac{\sum_{i \in \mathcal{I}}(\gamma'_i >0)}{|\mathcal{I}|} \times 100 \%, \label{eq:per_group_rate}
\end{equation}
where $(\gamma'_i >0)$ is $1$ if the inequality holds, and is $0$ otherwise. 
\subsubsection{The Hyperparameters}
The hyperparameters of interest are the number of replicas, the number of steps per episode, and the temperature schedule type. In particular, the notion of a ``step'' is equivalent to the local update. Table~\ref{table:pt_hyperparameters} shows the hyperparameter values used for the PTIC experiments, which were chosen such that all problems were solvable by each of the three algorithms. Of note, Ref.~\cite{rozada2019effects} shows competitive performance of the inverse-linear temperature distribution on hard Ising problems compared with other, more sophisticated temperature-setting methods. Some of these methods (e.g., the energy method in Ref.~\cite{hukushima1999domain}) do not necessarily work on the PTIC framework due to its not requiring MH updates to be performed at each step.

\begin{table}[H]
\centering
\caption{Numerical values or categorical types of the hyperparameters for the PTIC-WalkSAT algorithm}
\begin{tabular}{|c|c|}
\hline
\,\,\,\textbf{Hyperparameter}\,\,\, & \,\,\,\textbf{Value or type}\,\,\,\\
\hline
Number of replicas & 5 \\
Steps per episode & 5000 \\
Schedule & inverse-linear\\
Number of episodes & 1000\\
\hline
\end{tabular}
\label{table:pt_hyperparameters}
\end{table}

\subsubsection{The WalkSAT Heuristic Baseline}

The PTIC-WalkSAT algorithm was benchmarked against the WalkSAT algorithm, presented as Algorithm~\ref{algo:walksat_b}. Given that the PTIC-WalkSAT algorithm is naturally parallelized by its replicas, to allow a fair comparison, we parallelized the WalkSAT algorithm using multiple parallel runs instantiated with distinct random initial solutions. The number of parallel runs was set to the number of replicas in the PTIC-WalkSAT algorithm.

Table~\ref{table:pt_walksat_baseline} shows the hyperparameter values used for each parallel run. We used 0.5 as the random walk probability for all parallel runs, as it is the default setting of the WalkSAT codebase~\cite{walksatrepo}. The algorithm terminates as soon as one of the parallel runs finds a solution. The total number of iterations needed to find a solution is the number of iterations in the run that first finds the solution, multiplied by the number of parallel runs. 

\begin{table}[H]
\centering
\caption{Hyperparameter values for the WalkSAT algorithm baseline}
\begin{tabular}{|c|c|c}
\hline
\,\,\,\textbf{Hyperparameter}\,\,\, & \,\,\,\textbf{Value}\,\,\,\\
\hline
Random walk probability &  0.5 \\
\, Maximum number of iterations \, & \, 5,000,000 \, \\
\hline
\end{tabular}
\label{table:pt_walksat_baseline}
\end{table}

\subsubsection{The PA-WalkSAT Algorithm Baseline}

The second baseline is a parallelized version of the WalkSAT heuristic, or ``PA-WalkSAT'', which consists of multiple parallel runs of the WalkSAT heuristic, where the key difference is that the random walk probability parameters of each run are different, and match the ones on the replicas in the PTIC-WalkSAT algorithm. The algorithm can be viewed as a variant of the PTIC-WalkSAT algorithm that does not allow any replica exchanges and performs only one episode for each replica; for this reason, we set the maximum number of iterations to be 5,000,000, which corresponds to the number of steps per episode times the number of episodes of the values we use for benchmarking the PTIC-WalkSAT algorithm. The hyperparameter values are given in Table~\ref{table:pa_walksat_baseline}.
\begin{table}[H]
\centering
\caption{Hyperparameter values for the PA-WalkSAT algorithm baseline}
\begin{tabular}{|c|c|c}
\hline
\,\,\,\textbf{Hyperparameter}\,\,\, & \,\,\,\textbf{Value}\,\,\,\\
\hline
Number of replicas & 5 \\
\, Maximum number of iterations \, & \, 5,000,000 \, \\
\hline
\end{tabular}
\label{table:pa_walksat_baseline}
\end{table}
\subsubsection{Implementation and Computational Resources}
All of the benchmarked algorithms were fully implemented in-house with the use of the  \texttt{RUST} programming language. The results may be reproduced with other WalkSAT implementations, such as the one in Ref.~\cite{walksatrepo}. Regarding the computational resources, all of the experiments were performed on a Google Cloud Platform instance \mbox{(n2-standard-64)} located in the \mbox{us-central1-a zone.} The instance was equipped with an Intel Xeon CPU running at 2.80~GHz with 90~GB of RAM and 64 logical cores.

\section{Results}
\label{sec:results}

\subsection{Success Rates}

All 94 validation instances were solved with the use ofthe PTIC-WalkSAT algorithm, the WalkSAT heuristic, and the PA-WalkSAT algorithm. For each instance, each of the three algorithms attempted 1000 repeats, and the success rate for each is defined as the percentage of times the solution was found. Table~\ref{table:pt_success_rates} shows the median success rates of each algorithm for each group of problem instances. The PTIC-WalkSAT and PA-WalkSAT algorithms have higher success rates than the \mbox{WalkSAT} heuristic for all four groups; the \mbox{PTIC-WalkSAT} algorithm achieves higher success rates than the PA-WalkSAT algorithm for three of the groups. 
\begin{table}[H]
\centering
\caption{Success rates per problem group for each of the three algorithms}
\begin{tabular}{|c|c|c|c|c|}
\hline
 & \,\,\,Group 1\,\,\, & \,\,\,Group 2\,\,\, & \,\,\,Group 3\,\,\, & \,\,\,Group 4\,\,\,\\ 
\hline
\, PTIC-WalkSAT \, & 100\% & 25.4\% & 75.7\% & 59.5\% \\
WalkSAT & 100\% & 65.3\% & 67.9\% & 36.3\% \\
PA-WalkSAT & 100\% & 27.5\% & 73.9\% & 58.9\% \\
\hline
\end{tabular}
\label{table:pt_success_rates}
\end{table}
\subsection{Comparison of Problem Instances per Difficulty Level}

\begin{figure}[H]
    \centering
\includegraphics[width=8.5cm]{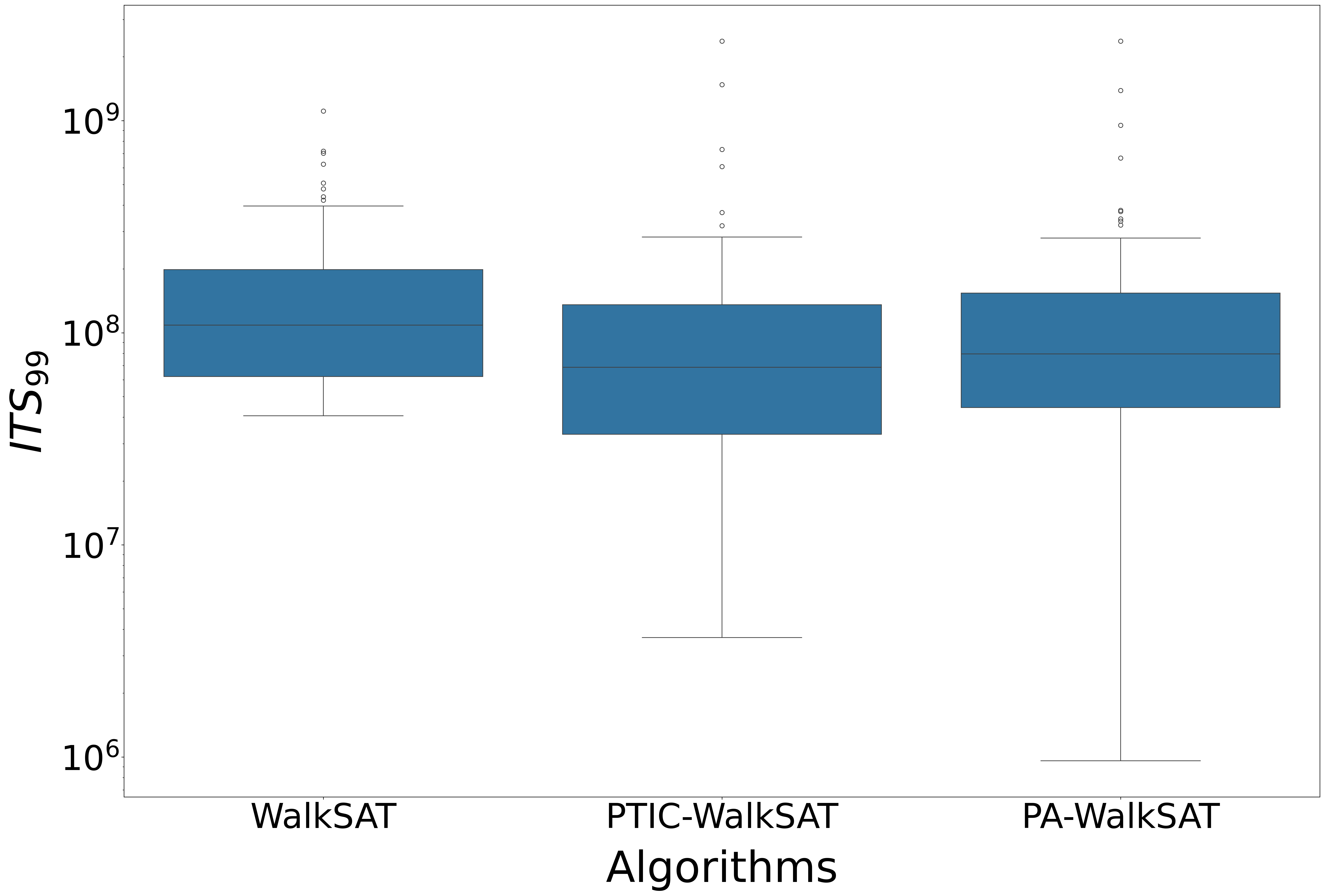}
    \caption{Box plots of the $ITS_{99}$ values for the three algorithms}
    \label{fig:pt_box_plot}
\end{figure}

To conduct a comprehensive assessment of the algorithms, we compare their performance at three difficulty levels, represented by percentiles of the validation instances: the 25th, 50th, and 75th percentiles, corresponding to easy, medium, and difficult cases, respectively. A visualization of this comparison is shown in Fig.~\ref{fig:pt_box_plot}, a box plot that illustrates both the percentiles and the spread of the data.

\begin{figure}[H]
   \centering
\includegraphics[width=8.5cm]{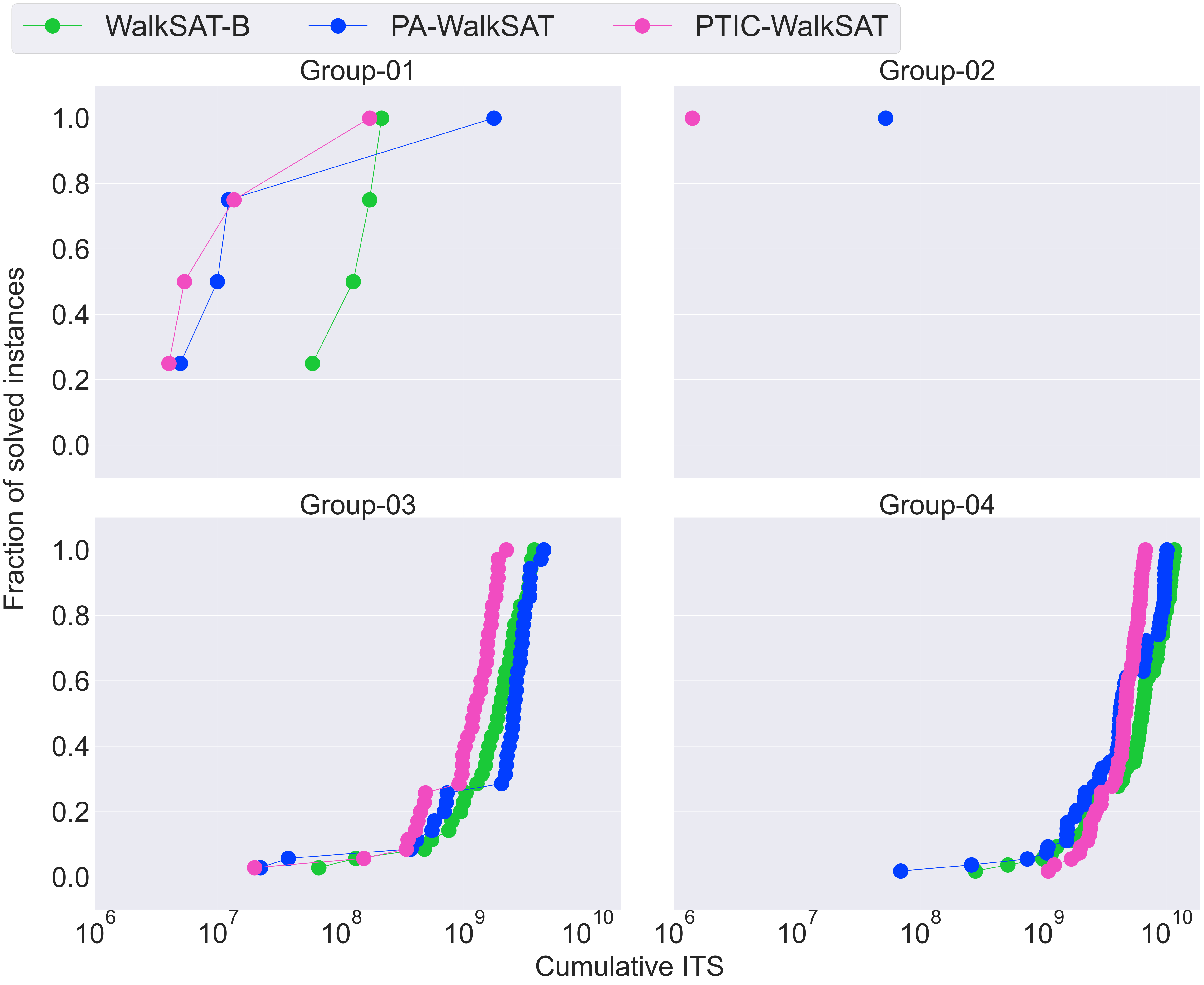}
   \caption{Groupwise cumulative $ITS_{99}$ plots}
   \label{fig:pt_cumulative_plot}
\end{figure}

As Fig.~\ref{fig:pt_box_plot} shows, the PTIC-WalkSAT and \mbox{PA-WalkSAT} algorithms both significantly outperform the WalkSAT heuristic. Table~\ref{tab:percentiles_its} shows the per-percentile $ITS_{99}$ values for the algorithms. We use the percentage difference metric defined in Sec.~\ref{sec:metric} to compare two algorithms. For the easiest 25\% of problem instances (the 25th percentile), the PTIC-WalkSAT algorithm outperforms the other two. The $ITS_{99}$ of the WalkSAT heuristic is 97.80\% higher than that of the PTIC-WalkSAT algorithm and 54.37\% higher than that of the PA-WalkSAT algorithm. For medium-difficulty instances (the 50th percentile), the $ITS_{99}$ of the WalkSAT heuristic is 62.39\% higher than that of the PTIC-WalkSAT algorithm. The PTIC-WalkSAT algorithm requires 4.59\% more iterations to find the solution than does the PA-WalkSAT algorithm. Regarding the 75th percentile, the WalkSAT heuristic requires 39.74\% more iterations than the \mbox{PA-WalkSAT} and PTIC-WalkSAT heuristics. 
\begin{table}[H]
\caption{$ITS_{99}$ values of the algorithms for three percentiles}
\label{tab:percentiles_its}
    \centering
    \begin{tabular}{|c|c|c|c|}
\hline
 Percentile & \,\,\,WalkSAT\,\,\, & \,\,\,PA-WalkSAT\,\,\, & \,\,\,PTIC-WalkSAT\,\,\,\\ 
\hline

25th & 9.0e+07	& 5.83e+07 & 4.55e+07 \\
\hline
50th & 1.77e+08 & 1.09e+08 & 1.14e+08\\
\hline
75th & 3.20e+08	& 2.29e+08 & 2.29e+08\\
\hline
\end{tabular}
\end{table}
The term ``cumulative $ITS_{99}$'' refers to the sum of the iterations needed to solve a given list of problem instances in some order.
The per-group cumulative plots are shown in Fig.~\ref{fig:pt_cumulative_plot}. The y-axis of each subfigure represents the fraction of solved instances, while the x value of any point shows the cumulative $ITS_{99}$ for all solved instances up to and including the instance associated with that point. Note that the plots depend on the ordering of the instance, but for each subfigure, the ordering is the same for all the algorithms. 

Figure~\ref{fig:pt_cumulative_plot} shows that, for random high-degree SAT problems, such as the ones in groups 2, 3, and 4, the \mbox{PTIC-WalkSAT} algorithm outperforms the two baselines, while its advantage becomes less consistent for SAT problems in group 1, which consists of structured problems. Note that there are only three instances from group 1 and one instance from group 2, which makes the corresponding cumulative plots sparse.

\subsection{Percentage of Instances with Improvement over Baselines}

\begin{figure}[H]
    \centering
\includegraphics[width=8.5cm]{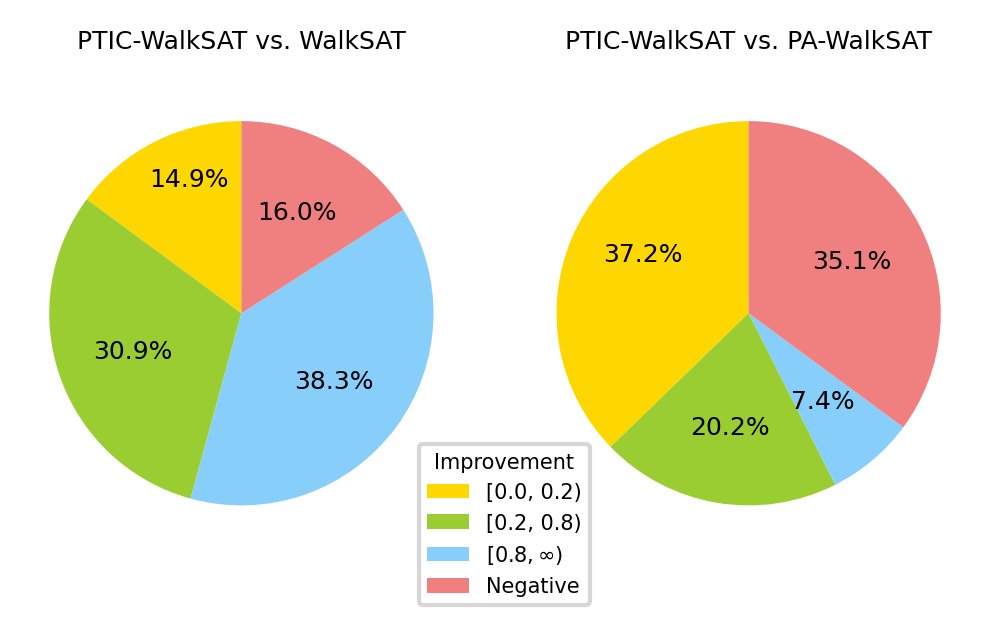}
    \caption{Percentage of instances that fall into each improvement range}
    \label{fig:pt_pie_plot}
\end{figure}

We use the percentage difference metric $\delta$ as defined in Sec.~\ref{sec:metric} to evaluate the improvement introduced by the PTIC-WalkSAT algorithm with respect to the baselines. The results are visualized in Fig.~\ref{fig:pt_pie_plot}. The charts show a breakdown of the percentage of the problem instances. Each coloured segment of a chart is annotated with a percentage indicating the proportion of the problem instances over which a corresponding improvement was obtained by the PTIC-WalkSAT algorithm when compared against a baseline. 

The pie chart on the left in Fig.~\ref{fig:pt_pie_plot} shows that for 38.3\% of the 94 problem instances, the \mbox{PTIC-WalkSAT} algorithm achieves a ``significant improvement'' (see Sec.~\ref{sec:metric}) over the WalkSAT heuristic in terms of $ITS_{99}$. Overall, the PTIC-WalkSAT algorithm performs better for 84\% of the instances, while the \mbox{WalkSAT} heuristic performs better for the other 16\%. The pie chart on the right shows significant improvement by the PTIC-WalkSAT algorithm over the \mbox{PA-WalkSAT} algorithm for 7.4\% of the problem instances. In total, the PTIC-WalkSAT algorithm outperforms the \mbox{PA-WalkSAT} algorithm for 64.9\% of the instances.
\subsection{A Per-Instance Perspective}

\begin{figure}[H]
    \centering
\includegraphics[width=8.5cm]{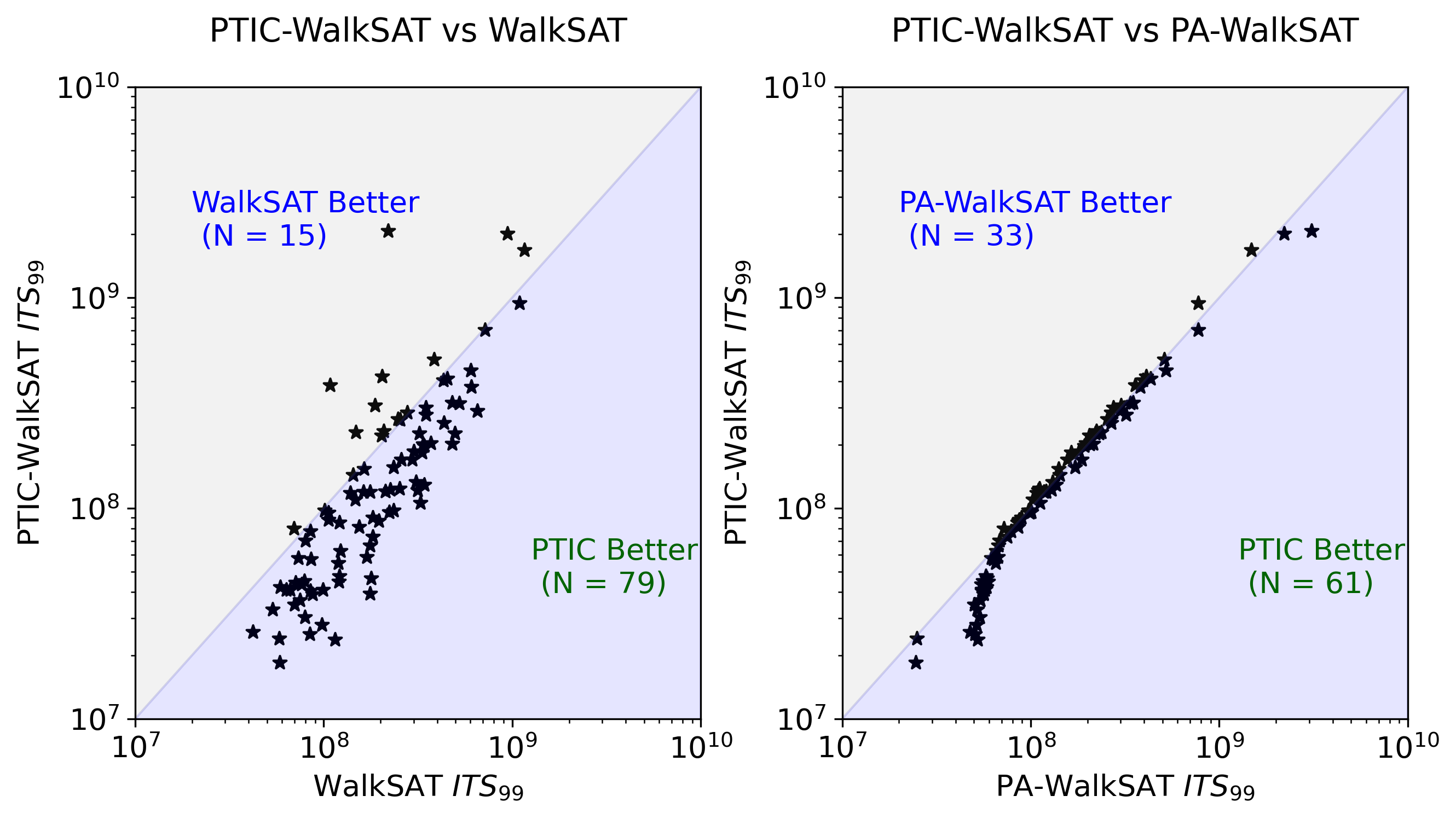}
    \caption{Correlation plots for $ITS_{99}$ values between the \mbox{PTIC-WalkSAT} algorithm and the baselines}
    \label{fig:pt_corr_plot}
\end{figure}

To provide a full per-instance view, correlation plots are presented in Fig.~\ref{fig:pt_corr_plot}. The y-axis shows the $ITS_{99}$ values of the PTIC-WalkSAT algorithm, while the x-axis shows the $ITS_{99}$ values of the baselines. Each point in a figure represents one of the 94 problem instances. The diagonal line distinguishes where the baseline performs better (shown in the upper triangular region) and where the PTIC-WalkSAT algorithm performs better (lower region), annotated with the total number of instances in each region.

The figure on the left provides a comparison between the WalkSAT heuristic and the \mbox{PTIC-WalkSAT} algorithm, where, for 79 of the 94 instances, the \mbox{PTIC-WalkSAT} algorithm outperforms the WalkSAT heuristic. The figure on the right provides a comparison of the PA-WalkSAT and PTIC-WalkSAT algorithms. The PTIC-WalkSAT algorithm performs better for 61 of the 94 instances.
\section{Discussion}
\label{sec:discussion}

Our work provides strong evidence that the physics-inspired PTIC framework can help build faster and more-efficient binary optimization solvers. Moreover, we estimate the energy consumption overhead for two candidate IMC hardware architectures and show that, in both cases, it would amount to a less than 1\% increase in energy consumption, making the framework an excellent fit for novel IMC solvers.

Our computational results show that the PTIC framework, when combined with the WalkSAT heuristic, outperforms parallelized versions of the WalkSAT heuristic. A significant proportion of the improvement comes from our treating the random walk probability parameter in the WalkSAT algorithm as the temperature parameter in the PT framework: high values correspond to exploratory regimes, and parameter values close to zero correspond to greedy exploitation regimes. Enabling a replica-exchange mechanism results in increased gains, albeit to a lesser extent.

We expect that a hardware implementation of the PTIC-WalkSAT algorithm will be faster and require less energy than standard and parallelized versions of the WalkSAT heuristic, while being less complicated to implement than other cooperative heuristics.

 The PTIC-WalkSAT algorithm has fewer total $ITS_{99}$ for 84.0\% of the problem instances than a standard \mbox{WalkSAT} heuristic and 64.9\% of the problem instances than the PA-WalkSAT algorithm, respectively. All three algorithms have the same number of replicas, and both the PTIC-WalkSAT algorithm and the parallelized \mbox{WalkSAT} heuristic share the same distribution of random walk probabilities, raising the question of why the \mbox{PTIC-WalkSAT} algorithm yields better results. We believe the answer lies in the replica-exchange mechanism. As discussed in Sec.~\ref{sec:bg_pre}, through the replica-exchange mechanism, high-temperature replicas have a better chance of exploring new solution basins and moving those solutions to lower temperatures so that the low-temperature replicas can efficiently converge to optimal solutions. Thus, it is not surprising that the \mbox{PTIC-WalkSAT} algorithm outperforms the PA-WalkSAT algorithm, which does not use a replica-exchange mechanism. For a more detailed overview of the replica-exchange mechanism, see Appendix~\ref{apx:internal_dynamics_ptic}.

This study has some limitations. First, we considered only replicas that implement the original WalkSAT heuristic formulation. In the years since it was first proposed, there have been many improvements to it. We believe the PTIC framework can be applied to most of these improved heuristics, as well as to other, more general SAT heuristics and to other problems by one choosing an appropriate solver parameter to represent the temperature of the replica. Second, we evaluated the PTIC-WalkSAT algorithm on a limited number of problem instances: we chose four SAT classes that are known to produce hard instances and filtered out the 83\% instances that were easiest to solve. It is possible that the PTIC-WalkSAT algorithm's ability to solve hard problem instances does not imply there will be comparable performance gains for easier problems. However, energy and runtime costs are dominated by hard problem instances, so we believe this is the area where improvements to solvers can have the greatest impact. Third, the PTIC framework generalizes PT and does not require MH updates; for this reason, its system dynamics may differ from the established understanding of PT~\cite{hukushima1996exchange}. A potential future research direction would be to generalize PT-specific techniques, such as temperature-setting methods under a generalized notion of what temperature represents in each replica. This area of research could unlock increased improvements from the replica-exchange mechanism.

\section{Conclusion}
\label{sec:conclusion}
Our study has been motivated by the recent growing interest in IMC binary optimization solvers. The principal idea of the study has been to propose an algorithmic framework to parallelize almost any IMC solver for better performance while controlling the overhead involved in the hardware implementation. We were inspired by the PT method and proposed the ``PT-inspired cooperative'' framework, which essentially launches multiple replicas of an IMC solver with different noise configurations. Following a scheme similar to that of PT, replicas are executed in a cooperative way. We applied the framework to the IMC solver proposed in Ref.~\cite{pedretti2023zeroth} to solve the SAT problems that the IMC solver struggles to handle. 

By conducting computational experiments, we have shown that the resultant PTIC-WalkSAT algorithm outperforms two baselines: the WalkSAT heuristic (a proxy for the IMC solver from Ref.~\cite{pedretti2023zeroth}) and the native parallelization of the WalkSAT heuristic, or ``PA-WalkSAT''. This shows the effectiveness of the PTIC framework, which we believe can be extended to other IMC solvers to achieve better performance.

\section*{Acknowledgements}
We thank our editor, Marko Bucyk, for his careful review and editing of the manuscript. We thank Gizem \"Ozbaygın for many useful discussions.

This material is based upon work supported by the Defense Advanced Research Projects Agency
(DARPA) through Air Force Research Laboratory Agreement No. FA8650-23-3-7313. The views,
opinions, and/or findings expressed are those of the authors and should not be interpreted as
representing the official views or policies of the Department of Defense or the U.S. Government.

\newpage
\section*{Appendix}

\setcounter{section}{0}
\setcounter{equation}{0}
\setcounter{figure}{0}
\setcounter{table}{0}
\setcounter{page}{1}
\setcounter{algorithm}{0}

\makeatletter

\renewcommand{\theequation}{A\arabic{equation}}
\renewcommand{\thefigure}{A\arabic{figure}}
\renewcommand{\thetable}{A\arabic{table}}
\renewcommand{\thesection}{A\arabic{section}}
\renewcommand{\thealgorithm}{A\arabic{algorithm}}

\renewcommand{\bibnumfmt}[1]{[#1]}
\renewcommand{\citenumfont}[1]{#1}

\section{Internal Dynamics of the PTIC-WalkSAT Algorithm}
\label{apx:internal_dynamics_ptic}

For a parallel tempering algorithm, it is important to know how replicas traverse different temperatures~\cite{earl2005parallel}. Having a replica frequently switch between two temperatures is not a desirable situation, as it would imply the replica exploring highly similar configuration spaces associated with the two temperatures. On the other hand, having a replica rarely travel between two temperatures is not desirable, either, as it would undermine the advantage of overcoming local energy traps by replica exchanges. 

To showcase the per-replica behaviours, we use the problem instance ``7-SAT-60'' as an example. Among 100 repeats of solving the chosen problem instance, we select the result from one of the repeats that successfully found the solution. 

\begin{figure}[H]
    \centering
\includegraphics[width=8.5cm]{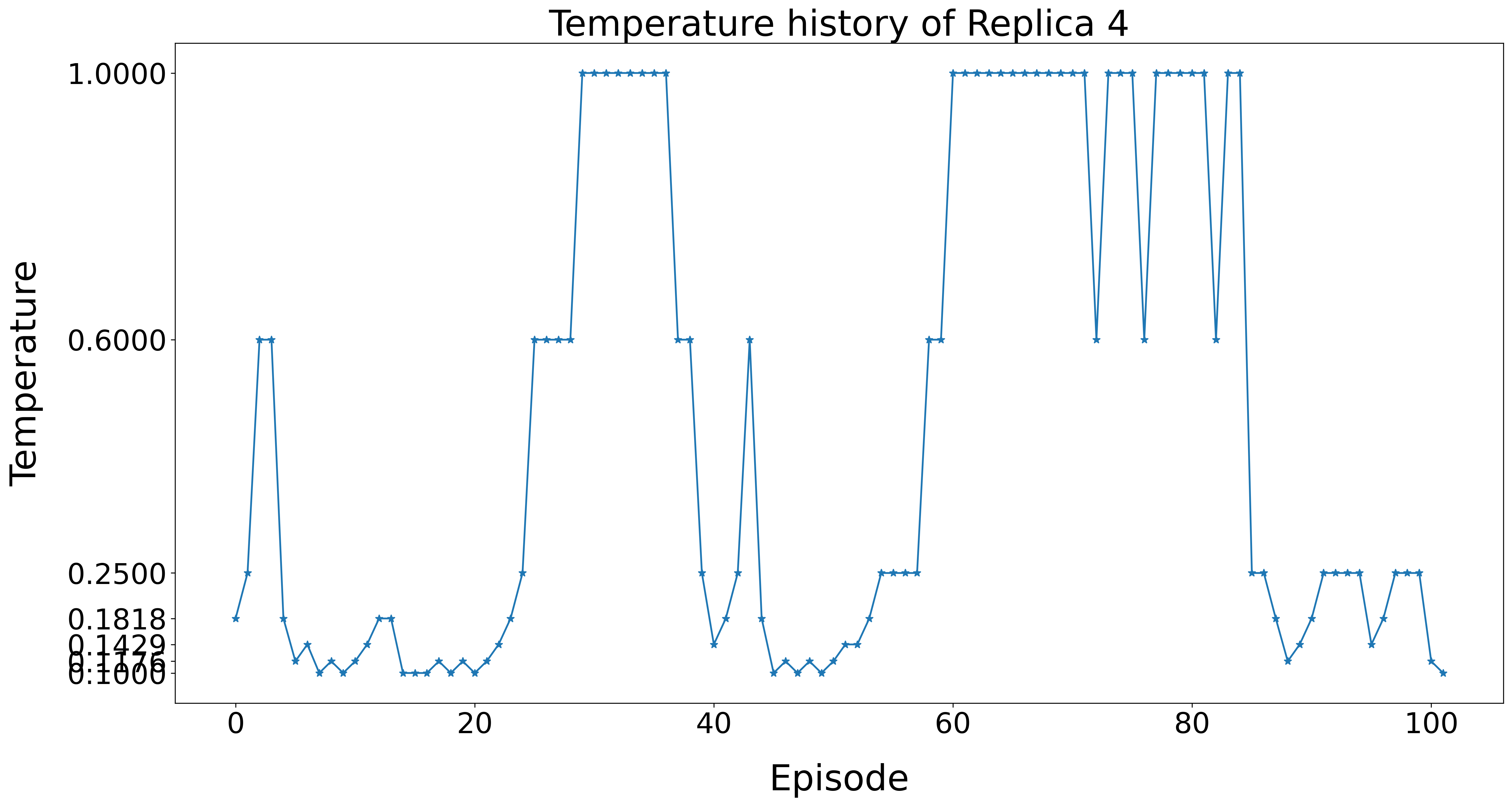}
    \caption{Replica 4 traversal history}
    \label{fig:replica_traveling_history}
\end{figure}

Figure~\ref{fig:replica_traveling_history} illustrates the traversal history across a full range of temperatures of a replica (``replica 4'') able to find the solution in the end. The figure assists in verifying that all the temperatures contribute to the path towards the solution. As is evident, the replica explores the solution space using relatively low temperatures within the first few episodes. Then, after approximately 20 episodes, it jumps to the highest temperature and drops to lower temperatures a few times. Eventually, the replica remains at lower temperatures and finds the solution using the lowest temperature. 
This process shows a typical search of a successful replica. It quickly finds a good lower-energy configuration in the beginning using lower temperatures; to escape from local minima, it jumps to higher temperatures to explore a much larger configuration space; once it finds a promising space in terms of its search for the global minimum, it performs a more refined search within the space and eventually the replica finds the lowest-energy configuration.

Aside from studying the traversal history of replicas, it can also be of benefit to learn how the energy evolves over the episodes for each replica, the main purpose of which is to verify whether replica exchanges carry replicas towards different energy landscapes. 

\begin{figure}[H]
   \centering
\includegraphics[width=8.5cm]{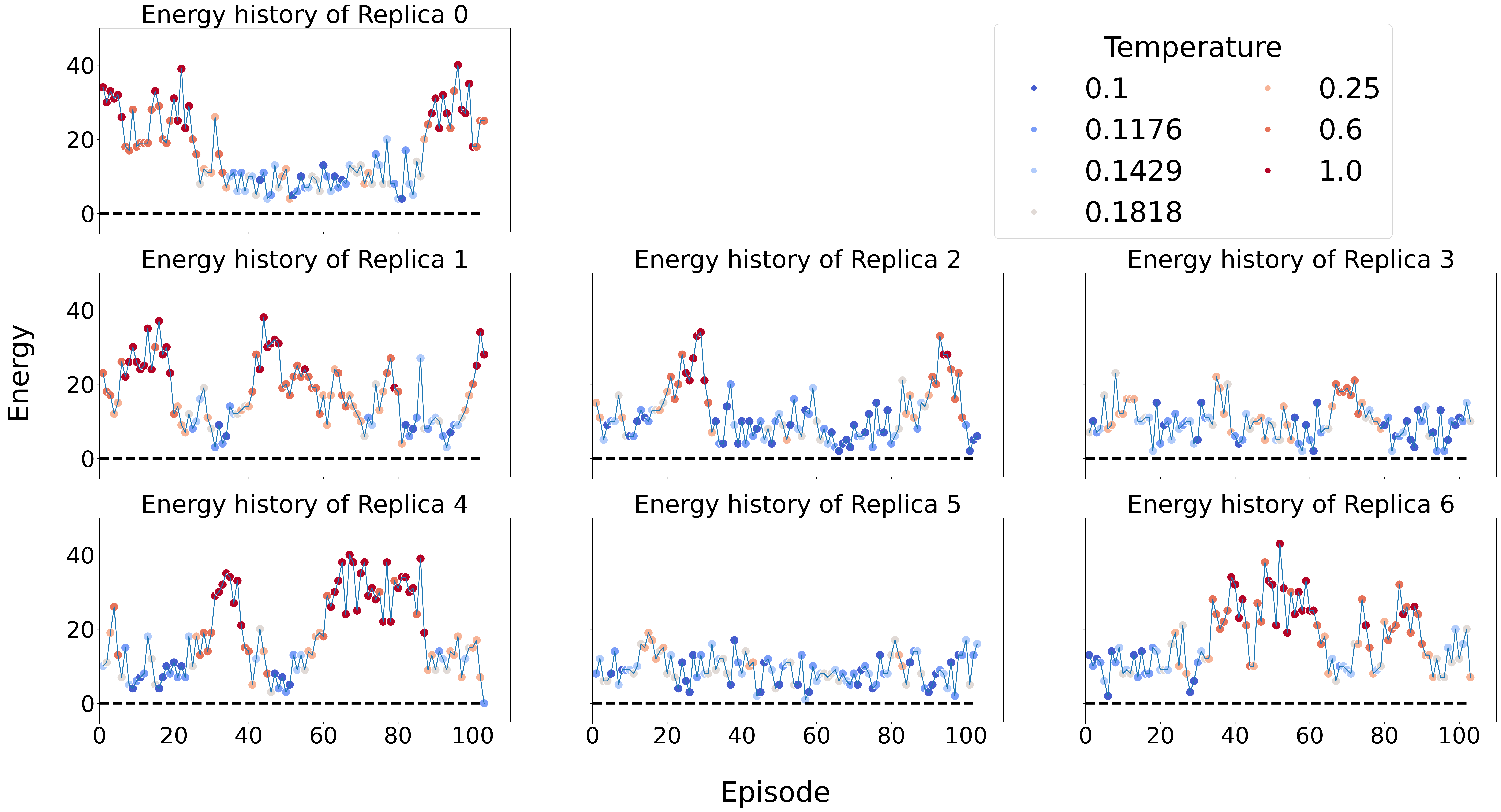}
   \caption{Per-replica energy history for seven replicas}
   \label{fig:per_replica_energy_history}
\end{figure}

\begin{figure}[H]
   \centering
\includegraphics[width=8.5cm]{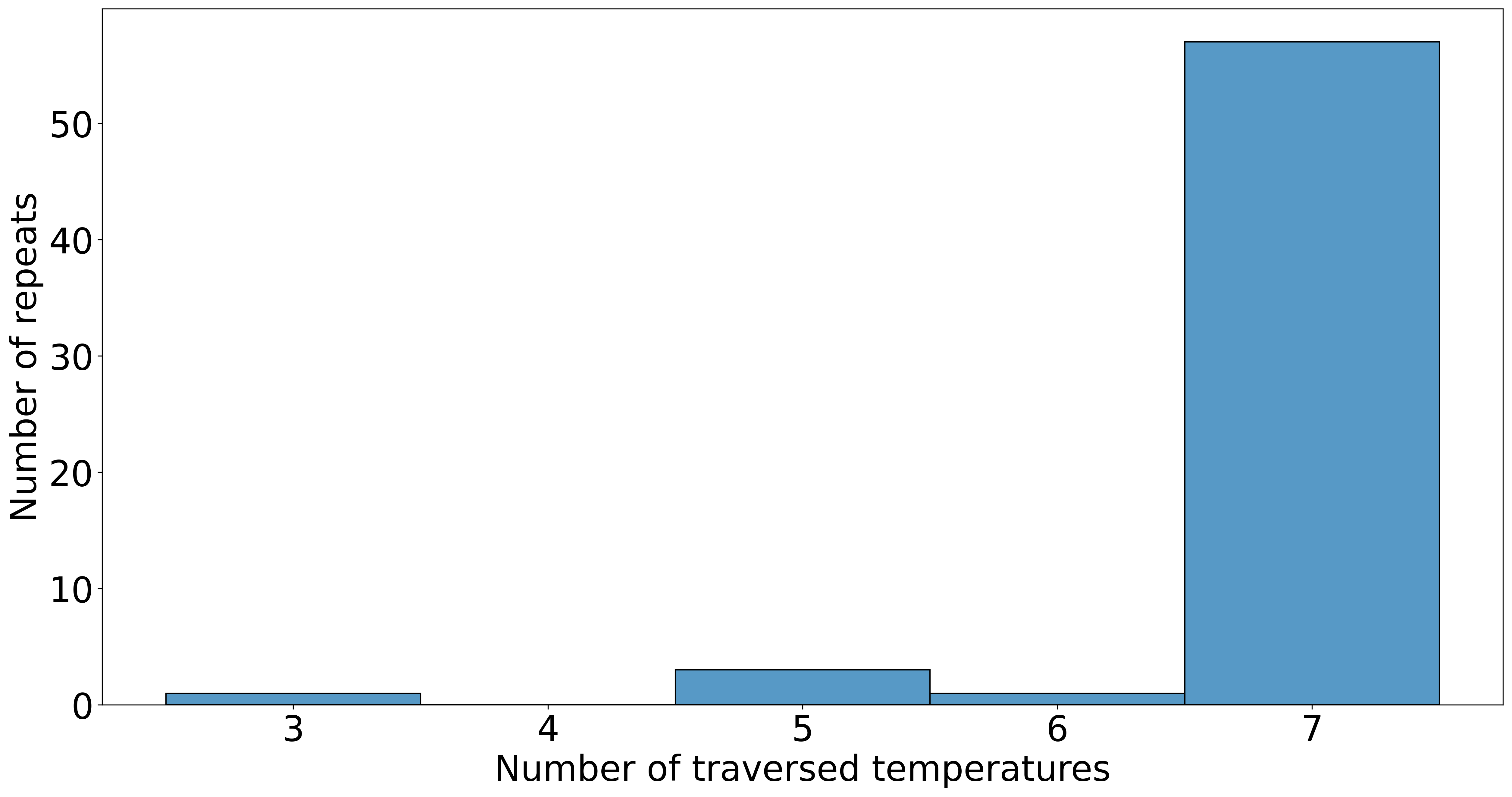}
   \caption{Number of traversed temperatures across 100 repeats}
   \label{fig:nb_temp_histogram}
\end{figure}

Figure~\ref{fig:per_replica_energy_history} shows the per-replica energy history for seven replicas. The colour scheme reflects the temperature at which a replica is executed during an episode.
It is insightful to compare the energy history of replica 5 with that of replica 4. Replica 5, which operates at lower temperatures, remains at lower energy levels throughout the process. In contrast, replica 4 alternates between the highest and lowest temperatures, covering a broader range of energy levels. This highlights the advantages of exploring the search space at higher temperatures, as replica 5 ultimately fails in finding a solution. The importance of using all temperature levels becomes even clearer when one examines the 100 repeats conducted for this problem instance. Figure~\ref{fig:nb_temp_histogram} shows a histogram of the number of temperatures traversed by the successful replicas over the 100 repeats. It is evident that most of the successful repeats need to traverse all the temperatures before finding a solution. 

\section{Full Benchmarking Results}
\label{apx:full_benchmarking_results}

Table \ref{tab:full_table_first_part} shows the full results for problem instances in groups 1, 2, and 3, and Table \ref{tab:full_table_second_part} presents the data for group 4. For both tables, the column headed ``instance'' gives the instance name, where the prefix indicates the group to which the instance is affiliated. The column headed ``WalkSAT ($R_{99}$)'' gives the mean $ITS_{99}$ and the uncertainty for the WalkSAT heuristic, while the value in parentheses is the number of repeats needed by the solver to find the solution with a  99\% probability of success~\ref{eq:its}. The columns headed ``PTIC-WalkSAT ($R_{99}$)'' and ``PA-WalkSAT ($R_{99}$)'' show the corresponding values for the PTIC-WalkSAT algorithm and the PA-WalkSAT algorithm, respectively.

\begin{table*}[htb]
\caption{\label{tab:full_table_first_part}Full experimental results for problem instances in group 1 (G1), group 2 (G2), and group 3 (G3). All the $ITS_{99}$ values and the uncertainties are expressed in millions of iterations. The  $R_{99}$ values represent the number of repeats needed to find the solution with 99\% certainty.}
\begin{tabular}{|c|c|c|c|}
\hline
Instance & PTIC-WalkSAT ($R_{99}$) & WalkSAT  ($R_{99}$) & PA-WalkSAT ($R_{99}$) \\
\hline
G1-015\_149{\_}181 & 3.7±0.1 (1.0) & 4.1±0.1 (1.0) & 22.7±0.5 (1.0) \\
G1-016\_149\_241 & 1.8±0.0 (1.0) & 3.1±0.1 (1.0) & 23.9±0.6 (1.0) \\
G1-016\_199\_229 & 7.4±0.2 (1.0) & 7.7±0.2 (1.0) & 23.0±0.6 (1.0) \\
G1-017\_191\_503 & 2078.6±570.3 (85.48) & 220.2±19.3 (8.83) & 3079.7±1056.6 (123.98) \\
G2-4-SAT-1 & 383.9±43.0 (15.7) & 107.8±7.6 (4.36) & 357.7±38.4 (14.31) \\
G3-6-SAT-1 & 24.1±0.6 (1.0) & 57.9±3.9 (2.77) & 24.8±0.6 (1.0) \\
G3-6-SAT-10 & 43.5±2.9 (2.77) & 76.2±5.1 (3.08) & 54.7±3.6 (2.77) \\
G3-6-SAT-100 & 285.1±27.7 (11.41) & 276.8±26.7 (11.18) & 267.9±25.4 (10.76) \\
G3-6-SAT-15 & 232.2±20.7 (9.31) & 207.7±17.7 (8.31) & 222.9±19.6 (8.92) \\
G3-6-SAT-17 & 42.2±2.8 (2.77) & 59.0±3.9 (2.77) & 58.2±3.9 (2.77) \\
G3-6-SAT-24 & 307.6±30.9 (12.4) & 187.2±15.4 (7.53) & 300.6±29.8 (12.02) \\
G3-6-SAT-25 & 40.9±2.7 (2.77) & 63.3±4.2 (2.77) & 57.1±3.8 (2.77) \\
G3-6-SAT-27 & 57.9±3.9 (2.77) & 73.1±4.9 (2.93) & 61.6±4.1 (2.77) \\
G3-6-SAT-31 & 422.5±49.0 (16.92) & 204.6±17.4 (8.26) & 409.0±46.7 (16.37) \\
G3-6-SAT-35 & 144.0±10.9 (5.78) & 143.0±10.8 (5.78) & 142.2±10.7 (5.7) \\
G3-6-SAT-37 & 57.2±3.8 (2.77) & 85.2±5.7 (3.44) & 63.4±4.2 (2.77) \\
G3-6-SAT-38 & 44.4±3.0 (2.77) & 71.1±4.7 (2.85) & 57.7±3.8 (2.77) \\
G3-6-SAT-4 & 156.7±12.1 (6.27) & 235.2±21.1 (9.43) & 171.8±13.7 (6.88) \\
G3-6-SAT-40 & 119.6±8.6 (4.8) & 162.3±12.7 (6.5) & 120.2±8.6 (4.81) \\
G3-6-SAT-41 & 109.8±7.7 (4.4) & 147.2±11.2 (5.91) & 102.7±7.1 (4.11) \\
G3-6-SAT-43 & 221.2±19.4 (8.87) & 202.1±17.1 (8.1) & 204.6±17.4 (8.22) \\
G3-6-SAT-47 & 30.4±2.0 (2.77) & 79.6±5.3 (3.21) & 53.5±3.6 (2.77) \\
G3-6-SAT-50 & 41.0±2.7 (2.77) & 66.3±4.4 (2.77) & 55.2±3.7 (2.77) \\
G3-6-SAT-54 & 97.4±6.7 (3.9) & 101.3±7.0 (4.06) & 99.7±6.9 (3.99) \\
G3-6-SAT-63 & 80.0±5.4 (3.21) & 69.5±4.6 (2.81) & 72.2±4.8 (2.89) \\
G3-6-SAT-66 & 33.0±2.2 (2.77) & 53.5±3.6 (2.77) & 52.1±3.5 (2.77) \\
G3-6-SAT-69 & 95.3±6.5 (3.82) & 105.9±7.4 (4.25) & 98.8±6.8 (3.96) \\
G3-6-SAT-7 & 299.5±29.7 (12.02) & 347.5±36.8 (13.93) & 274.4±26.3 (11.04) \\
G3-6-SAT-70 & 117.9±8.4 (4.72) & 138.3±10.3 (5.54) & 107.4±7.5 (4.31) \\
G3-6-SAT-72 & 34.8±2.3 (2.77) & 70.2±4.7 (2.83) & 50.3±3.3 (2.77) \\
G3-6-SAT-75 & 88.2±6.0 (3.53) & 105.7±7.4 (4.23) & 86.8±5.9 (3.49) \\
G3-6-SAT-78 & 25.9±1.7 (2.77) & 42.0±2.8 (2.77) & 47.5±3.2 (2.77) \\
G3-6-SAT-80 & 70.3±4.7 (2.83) & 79.8±5.3 (3.19) & 68.5±4.6 (2.77) \\
G3-6-SAT-82 & 169.9±13.5 (6.82) & 294.5±29.0 (11.78) & 156.7±12.1 (6.29) \\
G3-6-SAT-84 & 36.5±2.4 (2.77) & 75.0±5.0 (3.0) & 53.9±3.6 (2.77) \\
G3-6-SAT-86 & 263.5±24.8 (10.65) & 254.9±23.7 (10.26) & 266.1±25.1 (10.65) \\
G3-6-SAT-88 & 25.2±1.7 (2.77) & 84.3±5.7 (3.38) & 50.7±3.4 (2.77) \\
G3-6-SAT-90 & 81.4±5.5 (3.26) & 154.4±11.9 (6.18) & 85.6±5.8 (3.43) \\
G3-6-SAT-91 & 77.9±5.2 (3.14) & 85.2±5.7 (3.41) & 78.8±5.3 (3.15) \\
G3-6-SAT-97 & 229.1±20.3 (9.19) & 148.3±11.3 (5.94) & 237.2±21.3 (9.5) \\
\hline

\end{tabular}
\end{table*}

\begin{table*}[htb]
\caption{\label{tab:full_table_second_part}Full experimental results for problem instances in group 4 (G4). All the $ITS_{99}$ values and the uncertainties are expressed in millions of iterations. The  ($R_{99}$) values represent the number of repeats needed to find the solution with 99\% certainty.}
\begin{tabular}{|c|c|c|c|}
\hline
Instance & PTIC-WalkSAT  ($R_{99}$) & WalkSAT  ($R_{99}$) & PA-WalkSAT  ($R_{99}$)\\
\hline
G4-7-SAT-1 & 412.5±47.4 (16.6) & 452.7±54.2 (18.11) & 433.2±50.8 (17.34) \\
G4-7-SAT-100 & 95.7±6.5 (3.84) & 222.8±19.6 (8.91) & 100.6±6.9 (4.03) \\
G4-7-SAT-11 & 121.6±8.8 (4.88) & 315.1±32.0 (12.68) & 109.7±7.7 (4.4) \\
G4-7-SAT-12 & 58.8±3.9 (2.77) & 168.7±13.4 (6.75) & 67.0±4.5 (2.77) \\
G4-7-SAT-13 & 28.0±1.9 (2.77) & 98.1±6.7 (3.93) & 51.6±3.4 (2.77) \\
G4-7-SAT-14 & 315.7±32.1 (12.7) & 522.9±67.5 (21.17) & 338.0±35.4 (13.55) \\
G4-7-SAT-15 & 122.3±8.8 (4.92) & 226.2±20.0 (9.06) & 128.8±9.4 (5.15) \\
G4-7-SAT-16 & 449.8±53.7 (18.04) & 601.9±83.0 (24.17) & 522.0±67.0 (20.95) \\
G4-7-SAT-18 & 203.3±17.2 (8.14) & 369.8±40.4 (14.86) & 196.9±16.5 (7.89) \\
G4-7-SAT-19 & 45.1±3.0 (2.77) & 79.0±5.3 (3.19) & 56.0±3.7 (2.77) \\
G4-7-SAT-2 & 201.2±17.0 (8.06) & 335.9±35.1 (13.47) & 206.2±17.6 (8.27) \\
G4-7-SAT-20 & 700.5±104.5 (28.26) & 716.7±108.2 (28.86) & 771.0±120.5 (30.92) \\
G4-7-SAT-3 & 289.1±28.3 (11.67) & 653.6±93.8 (26.14) & 297.2±29.4 (11.89) \\
G4-7-SAT-30 & 405.6±46.4 (16.44) & 432.1±50.7 (17.32) & 393.5±44.1 (15.77) \\
G4-7-SAT-32 & 44.8±3.0 (2.77) & 120.3±8.6 (4.82) & 59.3±3.9 (2.77) \\
G4-7-SAT-33 & 129.4±9.5 (5.19) & 340.9±35.9 (13.7) & 135.9±10.1 (5.44) \\
G4-7-SAT-37 & 943.8±165.1 (38.27) & 1091.2±205.7 (43.94) & 772.2±121.2 (31.17) \\
G4-7-SAT-39 & 278.2±26.8 (11.16) & 349.0±37.1 (14.05) & 319.1±32.6 (12.85) \\
G4-7-SAT-4 & 265.3±25.0 (10.65) & 248.1±22.8 (10.02) & 254.4±23.6 (10.21) \\
G4-7-SAT-41 & 39.0±2.6 (2.77) & 87.3±5.9 (3.5) & 56.4±3.8 (2.77) \\
G4-7-SAT-43 & 120.5±8.7 (4.83) & 211.8±18.3 (8.49) & 120.4±8.6 (4.83) \\
G4-7-SAT-45 & 73.3±4.9 (2.95) & 181.6±14.8 (7.26) & 74.9±5.0 (3.0) \\
G4-7-SAT-46 & 40.9±2.7 (2.77) & 98.8±6.8 (3.98) & 55.5±3.7 (2.77) \\
G4-7-SAT-47 & 40.4±2.7 (2.77) & 85.1±5.7 (3.42) & 55.2±3.7 (2.77) \\
G4-7-SAT-49 & 124.2±9.0 (5.0) & 252.8±23.4 (10.12) & 110.9±7.8 (4.44) \\
G4-7-SAT-51 & 133.1±9.8 (5.35) & 309.3±31.1 (12.39) & 130.2±9.5 (5.21) \\
G4-7-SAT-52 & 55.1±3.7 (2.77) & 119.4±8.6 (4.78) & 65.2±4.3 (2.77) \\
G4-7-SAT-53 & 85.7±5.8 (3.43) & 121.0±8.7 (4.86) & 85.0±5.7 (3.4) \\
G4-7-SAT-56 & 86.9±5.9 (3.49) & 196.1±16.4 (7.88) & 86.4±5.8 (3.46) \\
G4-7-SAT-58 & 153.3±11.8 (6.17) & 163.9±12.9 (6.56) & 140.1±10.5 (5.61) \\
G4-7-SAT-59 & 46.5±3.1 (2.77) & 177.9±14.4 (7.12) & 58.5±3.9 (2.77) \\
G4-7-SAT-60 & 226.9±20.1 (9.1) & 497.0±62.3 (19.91) & 220.2±19.2 (8.81) \\
G4-7-SAT-61 & 1680.0±411.4 (70.27) & 1163.5±227.3 (46.91) & 1483.4±330.6 (59.54) \\
G4-7-SAT-62 & 66.4±4.4 (2.77) & 175.9±14.2 (7.05) & 67.4±4.5 (2.77) \\
G4-7-SAT-67 & 197.5±16.6 (7.92) & 341.8±36.0 (13.76) & 191.8±15.9 (7.68) \\
G4-7-SAT-68 & 97.4±6.7 (3.9) & 234.0±21.0 (9.43) & 97.2±6.7 (3.89) \\
G4-7-SAT-70 & 119.2±8.5 (4.77) & 176.6±14.2 (7.09) & 117.4±8.4 (4.7) \\
G4-7-SAT-71 & 39.4±2.6 (2.77) & 176.5±14.2 (7.07) & 56.1±3.7 (2.77) \\
G4-7-SAT-73 & 2007.7±538.6 (82.26) & 945.2±167.8 (39.33) & 2201.6±626.0 (90.8) \\
G4-7-SAT-74 & 227.2±20.1 (9.1) & 321.2±32.9 (12.89) & 230.9±20.6 (9.28) \\
G4-7-SAT-75 & 106.3±7.4 (4.27) & 325.2±33.5 (13.04) & 111.8±7.9 (4.48) \\
G4-7-SAT-79 & 184.1±15.1 (7.41) & 331.4±34.4 (13.25) & 164.2±12.9 (6.59) \\
G4-7-SAT-8 & 62.7±4.2 (2.77) & 122.9±8.9 (4.94) & 65.8±4.4 (2.77) \\
G4-7-SAT-80 & 508.7±64.6 (20.49) & 384.7±42.8 (15.47) & 509.1±64.5 (20.38) \\
G4-7-SAT-81 & 202.5±17.1 (8.11) & 480.9±59.3 (19.28) & 214.7±18.6 (8.59) \\
G4-7-SAT-82 & 254.0±23.5 (10.17) & 433.7±51.0 (17.4) & 266.9±25.2 (10.68) \\
G4-7-SAT-83 & 23.8±1.6 (2.77) & 114.4±8.1 (4.59) & 52.4±3.5 (2.77) \\
G4-7-SAT-84 & 47.6±3.2 (2.77) & 121.0±8.7 (4.84) & 57.8±3.8 (2.77) \\
G4-7-SAT-86 & 185.8±15.2 (7.45) & 301.0±29.9 (12.06) & 179.6±14.6 (7.19) \\
G4-7-SAT-9 & 377.0±41.5 (15.14) & 607.0±84.2 (24.48) & 381.5±42.2 (15.28) \\
G4-7-SAT-92 & 18.5±0.4 (1.0) & 58.5±3.9 (2.77) & 24.6±0.6 (1.0) \\
G4-7-SAT-95 & 170.4±13.6 (6.88) & 257.4±23.9 (10.3) & 186.3±15.3 (7.52) \\
G4-7-SAT-96 & 90.2±6.1 (3.62) & 181.8±14.8 (7.29) & 89.1±6.0 (3.58) \\
G4-7-SAT-98 & 316.3±32.2 (12.7) & 480.4±59.3 (19.28) & 349.8±37.1 (14.01) \\
\hline
\end{tabular}
\end{table*}

\clearpage

\bibliographystyle{ptsat_bib_style}
\bibliography{ptsat}

\end{document}